\documentclass{jetpl}
\twocolumn

\lat

\title{ SANC integrator in the progress: QCD and EW contributions}

\rtitle{ SANC integrator in the progress: QCD and EW  contributions}

\author{
D.\,Bardin$^{a}$\/\thanks{e-mail: sanc@jinr.ru},S.\,Bondarenko$^{b}$,P.\,Christova$^{a,}$,L.\,Kalinovskaya$^{a}$,L.\,Rumyantsev$^{a,c}$,A.\,Sapronov$^{a}$,W.\,von Schlippe$^{d}$.}

\rauthor{D.\,Bardin,S.\,Bondarenko,P.\,Christova,L.\,Kalinovskaya,L.\,Rumyantsev,A.\,Sapronov,W.\,von Schlippe.}

\sodauthor{Bardin, Bondarenko, Christova, Kalinovskaya, Rumyantsev, Sapronov, von Schlippe}

\address{
$^{a}$ Dzhelepov Laboratory for Nuclear Problems, JINR,      \\
        ul. Joliot-Curie 6, RU-141980 Dubna, Russia;         \\
$^{b}$ Bogoliubov Laboratory of  Theoretical Physics, JINR,  \\
        ul. Joliot-Curie 6, RU-141980 Dubna, Russia;         \\
$^{c}$  Rostov University, Russia;     \\
$^{d}$  Petersburg Nuclear Physics Institute, Gatchina, 188300, Russia.
}

\dates{17 July 2012}{*}

\abstract{
Modules and packages for the one-loop calculations at partonic level
represent the first level of SANC output computer product.
The next level represents Monte Carlo integrator {\tt mcsanc},
realizing fully differential hadron level calculations (convolution with PDF) for the HEP
processes at LHC.
In this paper we describe the implementation into the framework {\tt mcsanc}
first set of processes: DY NC, DY CC, $f_1 \bar{f}'_1 \to H W^\pm(Z)$ and single top production.
Both EW and QCD NLO corrections are taken into account.
A comparison of SANC results with those existing in the world literature is given.
}

\begin{document}

\maketitle

\section{Introduction}

Recent reviews of theoretical predictions and their uncertainties for basic LHC
processes in the Standard Model (SM) can be found in Reports~\cite{Dittmaier:2012vm}
and~\cite{Dittmaier:2011ti}.

The interpretation of high-quality data of the LHC demands an
equally high precision in the theoretical predictions at the level of quantum corrections.
Apart from a detailed knowledge of higher-order EW and QCD corrections,
the combination of their effects must be investigated.
Advanced computational tools were developed to control the interplay of EW and QCD corrections:
~\cite{Balossini:2009sa},~\cite{Denner:2011id},~\cite{Bernaciak:2012hj},~\cite{Barze:2012tt} and
\cite{Yost:2012mf}.

In this paper new results of the computer system SANC
(Support of Analytic and Numerical Calculations for experiments at Colliders)~\cite{Andonov:2004hi}
are presented. In this system  it is possible
to achieve the one loop level predictions in the EW and QCD sectors on the same platform of the analytic procedures.

The first level of the computer products SANC are: analytical modules for scalar
Form Factors (FF) and Helicity Amplitudes (HA)
and accompanying bremsstrahlung contributions (BR or MC) and
the {\tt s2n.f} package producing the FORTRAN codes~\cite{Andonov:2008ga}.

In this paper we discuss in some detail the results of the implementation at
hadronic level in the newly developed
{\tt mcsanc-1.0} integrator, based on the above mentioned modules. The processes are marked by process
identifiers: pid=cnn, c=charge: 0-NC, $\pm$-CC, and: nn=01($e$), 02($\mu$), 03($\tau$) etc, see below.

$\bullet$ Drell--Yan-like single $W$ production: pid $= \pm102$.
\begin{equation}
 \bar{d}  +  u  \to  l^+ + \nu_l
\label{eqDYCC}
\end{equation}

$\bullet$ Drell--Yan-like single $Z$ production: pid $= 002$.
\begin{equation}
q + \bar{q} \to l^+ +l^-
\label{eqDYNC}
\end{equation}

$\bullet$  $HW^{\pm}(Z)$ production: pid $=\pm 104$ ($004$).

At the parton level we consider
\begin{equation}
 f_1\bar{f}'_1HW^\pm(Z) \to 0
\label{eqWH}
\end{equation}
(where $f_1$ stands for a {\em massless fermion} of the SM,
while specifically for bosons we use $Z,\,W^{\pm},\,H$).
It should be emphasized also that the notation
$ffHW\to 0$ means that all external 4-momenta flow inwards; this is the standard SANC convention
which allows to compute one-loop covariant amplitude (CA) and form factors (FF) only once
and obtain CA for a specific channel by means of a crossing transformation.

$\bullet$ the $s$ and $t$ channels of single top quark production: pid $= \pm 105(s), \pm 106(t)$.
\begin{equation}
 t \bar{b} \bar{u} d\to 0~~\mbox{and}~~\bar{t} b u \bar{d} \to 0.
\label{eqst}
\end{equation}

Previous studies of these processes by SANC system, i.e. creation of the
analytic platform and modules at the parton level were presented in
\cite{Andonov:2004hi,Arbuzov:2005dd,Arbuzov:2007ke,Arbuzov:2007db,Andonov:2009nn,Bardin:2010mz,Bardin:2011ti}.

The paper is organized as follows.
First, an overview of the {\tt mcsanc-1.0} integrator is given in section~\ref{SectionInt}.
In subsection ~\ref{SubSectionDes} we describe
a list of contributions to hard sub-processes and  introduce their enumeration.
Then we describe the parallel calculations issue.
Section~3\, contains numerical results for the processes in QCD and EW sectors.
Input parameters, kinematical cuts, and the used PDFs can be found in subsection~\ref{SubSectionNumIn}.
Further, we present the complete predictions for inclusive cross sections at LO and NLO levels in the EW and QCD
sectors for processes~(\ref{eqDYCC})--(\ref{eqst}).
We systematically compare our results for NLO QCD corrections with the
program MCFM~\cite{Campbell:arXiv1007.3492},~\cite{MCFMhomepage} and,
whenever possible, with other codes existing in the literature:
~\cite{Ciccolini:2003jy},~\cite{Brein:2004ue} (EW) and~\cite{Harris:2002md} (QCD).

In section~4 we summarize our results.

\section{SANC Integrator}
\label{SectionInt}

\subsection {\bf Description of {\tt id's} for hard sub-processes}
\label{SubSectionDes}

At NLO level several hard sub-processes contribute to a given process. In general, it consists of several
parts: LO--lowest order, Virt--virtual, Real--real brems(glue)strahlung and Subt--subtraction;
Real, in turn, is subdivided into Soft and Hard contributions.
We enumerate them through {\tt id=0}--{\tt 6}:
\begin{itemize}
\item[id0:]~LO, 2$\to$ 2, tree-level, $q\bar{q'}$ NC or CC sub-processes.
\item[id1:]~Subt term, responsible for the subtraction of the initial quark mass ($m_q$) singularities for $q\bar{q'}$
           sub-processes, computed in a given subtraction scheme ($\overline{\mbox{MS}}$ or DIS). It depends on $\ln(m_q)$.
\item[id2:]~Virt represents only the NLO EW parts, stands for pure EW one-loop virtual contributions. It depends on $m_q$ and
           may depend on an infrared regulator (e.g. on an infinitesimal photon mass). It is not present for QCD NLO
           contributions, where it is added to the soft contribution (see next item).
           For DY NC NLO EW process it contains all virtual contributions, both EW and QED.
\item[id3:]~For all processes, except DY NC, this stands for the sum of Virt and Real Soft (QED/QCD) contributions, therefore
           it does not depend on the infrared regulator but depends on $m_q$ and on the soft-hard separator $\bar{\omega}$.
           For DY NC NLO EW processes it is just the Real Soft QED contribution that depends on the infrared regulator, on $m_q$
           and on the soft-hard separator $\bar{\omega}$.
\item[id4:]~For all processes this is just the Real Hard (QED/QCD) contribution that depends on $m_q$ and on $\bar{\omega}$.
\item[id5:]~Subt term is responsible for the subtraction of the initial quark mass
singularities for $gq(gq')$ sub-processes
           (also computed in $\overline{\mbox{MS}}$ or DIS schemes). It contains logarithmic singularities in $m_q$.
\item[id6:]~The gluon-induced sub-process--an analog of {\tt id4} for $gq(gq')$ sub-processes. They also contain logarithmic
           mass singularities which cancel those from {\tt id5}.

\end{itemize}

The quark mass is used to regularize the collinear divergences, the
soft-hard separator is a remainder of infrared divergences.
The sum of contributions with {\tt id3} and {\tt id4} is independent of $\bar{\omega}$.
The sums {\tt id1+id2+id3+id4} and {\tt id5+id6} are separately independent of $m_q$. Therefore, the entire NLO sub-process
is independent of both unphysical parameters $\bar{\omega}$ and $m_q$.

\subsection{Parallel calculations}
The {\tt mcsanc} program takes advantage of parallelization in the 
Cuba library~\cite{Hahn:2004fe},~\cite{Hahn:CUBAhp}, used as a Monte Carlo integrating tool. 
However, the parallelization efficiency is reduced by the overhead of inter-process communications.

\begin{figure}[h]
\includegraphics*[height=7.5cm,width=8cm]{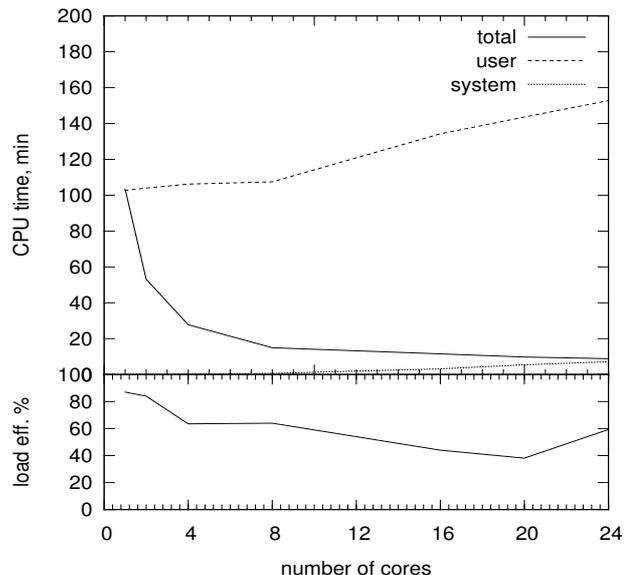}
\caption{Figure 1. CPU usage and load efficiency for the {\tt mcsanc} program depending
on the number of processor cores}
\label{fig:cpu-usage}
\end{figure}

Figure~\ref{fig:cpu-usage} shows time required to complete the NLO EW
cross section calculation depending on the number of active CPU cores. The
test was run on a dual-processor Intel~\textsuperscript{\textregistered}
Xeon~\textsuperscript{\textregistered} machine with 12 real (24 virtual)
cores with Linux operating system. The upper plot summarizes multicore
CPU productivity: ``total'' is the
wall clock time passed during the run; ``user'' is the CPU time consumed by the
program (roughly equals to wall clock time multiplied by the number of cores in
case of 100\% efficiency); ``system'' is the time spent by the operating system
on multiprocessing service.  

One can see that the parallelization is
efficient with number of cores up to 8, after which the total run time does not
significantly decrease and the overhead CPU time (``user'') grows.  It is also apparent from
the lower plot that the average CPU load efficiency drops below 50\%
with more than 8 cores active.

\section{Numerical results}
\label{SectionNum}

In this section the results, obtained by the {\tt mcsanc} integrator, realizing fully differential hadron level
calculations for the processes (\ref{eqDYCC})--(\ref{eqst}) are presented.

We produce comparison with numerical results for the NLO QCD corrections for all our processes between
SANC and \cite{Campbell:arXiv1007.3492}.
For the NLO electroweak corrections
for Drell--Yan NC and CC processes (pid $= 002,\pm 102$) this was done early within the workshop~\cite{Buttar:2008jx}.
For $WH$ production, i.e. pid $= \pm 104$, in EW sector
between SANC and \cite{Ciccolini:2003jy} and in QCD sector between SANC and \cite{Harris:2002md}
for pid $= \pm 105,\pm 106$.

\subsection{Setup: PDF, cuts, input parameters}
\label{SubSectionNumIn}
For the numerical results in this section we have used the following setup.

$\bullet$ PDF set, scales, $\alpha_s$. We use {\tt CT10(f[scale])} PDF from the LHAPDF library and compute $\alpha_s$
via a call to {\tt alphasPDF(r[scale])}. Usually we set factorization scale ({\tt fscale}) equal to renormalization
scale ({\tt rscale}) and different for the processes under consideration: $M_V$ for DY-like single $V$ production;
$M_{V+H}$ for the processes Eq.(\ref{eqWH}); $m_{t}$ for the processes Eq.(\ref{eqst}).

$\bullet$ Phase-space cuts. We use loose cuts: for the final state particle transverse momenta $p_{T}\geq 0.1\,$GeV,
no cuts for their rapidities and for the neutral current DY, in addition, $M_{l^+l^-}\geq 20\,$GeV.
We demonstrate numerical results for Drell--Yan only for muon case and we are not dealing with effects of
recombination. We choose $\omega = 10^{-4}$ and the cms energy $\sqrt{s_0}=14\,$TeV if not stated otherwise.

$\bullet$ Set of EW scheme and input parameters. We choose the $G_{\mu}$ EW scheme, and input parameters are taken
from PDG-2011 (on 16/05/2012):\\
Coupling constants:
 $\alpha =1/137.035999679$, $G_{F} = 1.16637\times 10^{-5}$.
Boson masses: $M_{W}= 80.399$\,GeV, $M_{Z}= 91.1876$\,GeV, $M_{H}= 120$\,GeV.
Boson widths: $\Gamma_Z$ = 2.4952\,GeV, $\Gamma_W$ = 2.085\,GeV.
CKM matrix: $V_{ud} = 0.9738$, $V_{us}= 0.2272$, $V_{cd}= 0.2271$, $V_{cs}= 0.9730$.\\
Lepton masses: $m_e = 0.510998910$\,MeV, $m_\mu = 0.105658367$\,GeV,
$m_\tau = 1.77682$\,GeV. Heavy quark masses: $ m_b = 4.67 $\,GeV, $m_t = 172.9$\,GeV.
Masses of the four light quarks are taken from~\cite{Dittmaier:2009cr}:
 $ m_d = 0.066$\,GeV,
 $ m_u = 0.066$\,GeV,
 $ m_s = 0.150$\,GeV,
 $ m_c = 1.2  $\,GeV.


\subsection{Example of $M_{\mu^+\mu^-}$ distributions, DY NC}

The standard ATLAS Monte Carlo (MC) generation uses the PYTHIA--PHOTOS chain of programs.
PYTHIA~\cite{Sjostrand:2006za},\cite{Sjostrand:2007gs}
has the leading order (LO) matrix element for a given process and takes into account
Parton Showers (PS);
PHOTOS~\cite{Barberio:1990ms},~\cite{Golonka:2005pn} and~\cite{Was:2008zu} describes multiphoton emission
from the Final State (FSR) di-lepton system.
This procedure does not include certain next-to-leading order (NLO) EW corrections, like Pure Weak (PW)
contributions, Initial--Final QED interference (IFI) and what remains from Initial State Radiation
(ISR) after subtraction of collinear divergences. In SANC one can evaluate the entire
effect of these corrections as a difference of complete NLO EW corrections and the QED FSR corrections.
See, for example, the distribution of the complete NLO EW corrections over $Z$ boson
invariant mass, $\delta(M_{\mu^+\mu^-})$, for Drell--Yan-like single $Z$ production around $Z$ resonance,
($\delta=(d\sigma^{\rm NLO}/d M_{\mu^+\mu^-})/(d\sigma^{\rm LO}/d M_{\mu^+\mu^-})-1$
and with taking into account only QED FSR corrections, Figure~2.
\begin{figure}[!h]
\begin{center}
\includegraphics[width=8.75cm]{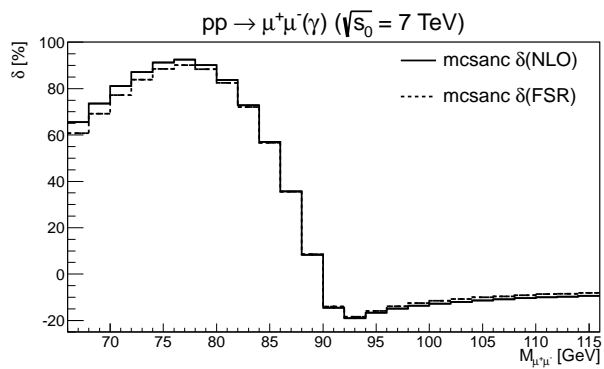}
\caption{Figure 2.~$\delta$ in $\%$ with complete NLO EW (solid histogram) and FSR (dashed histogram)
corrections \label{miv_nlo}}
\end{center}
\end{figure}

\begin{figure}[!h]
\begin{center}
\includegraphics[width=8.75cm]{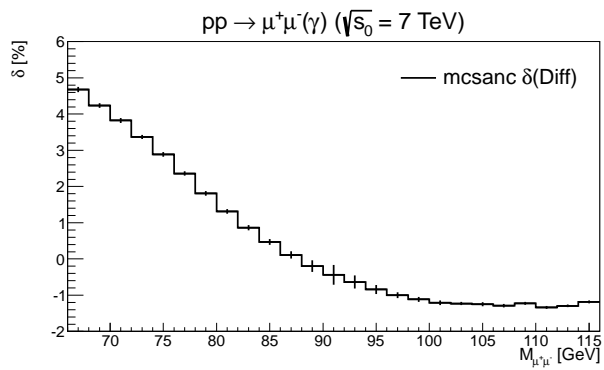}
\caption{Figure 3. Difference correction $\delta^{\rm Diff}$ in $\%$ for the distribution over $M_{\mu^+\mu^-}$
\label{miv_mis}}
\end{center}
\end{figure}


NLO and FSR distributions in Figure 2 are barely distinguishable.
The difference $\delta^{\rm Diff}=\delta^{\rm NLO - FSR}(M_{\mu^+\mu^-})$  is shown in Figure 3.

As is seen, for the $M_{\mu^+\mu^-}$ interval around the $Z$ resonance, $\delta^{\rm Diff}$ varies from $+5\%$
at the lower edge to $-1\%$ at the upper edge and therefore cannot be neglected, if the precision tag
is equal to 1$\%$, say.

\subsection{Numerical results and comparison for LO, NLO EW, NLO QCD RC}
\label{SubSectionNumComp}

$\bullet$ {In Tables 1--3 we present LO and NLO inclusive cross sections for processes (1--2), (3) and (4),
respectively.
\begin{table}[!h]
\begin{center}
\begin{tabular}{|l|r|r|r|}
\hline
 pid             &   002       &   102        &   -102      \\ \hline 
 LO              &   3338(1)   &   10696(1)   &   7981(1)   \\
 LO MCFM         &   3338(1)   &   10696(1)   &   7981(1)   \\ 
 NLO QCD         &   3388(2)   &   12263(4)   &   9045(4)   \\ 
 NLO MCFM        &   3382(1)   &   12260(1)   &   9041(5)   \\ 
$\delta_{QCD}$    &   1.49(3)   &   14.66(1)   &   13.35(3)  \\
 NLO EW          &   3345(1)   &   10564(1)   &   7861(1)   \\  
$\delta_{EW}$     &   0.22(1)   &   -1.23(1)   &   -1.49(1)  \\  \hline
\end{tabular}
\end{center}
\caption{Table 1: NC and CC DY processes, i.e. for pid $ = 002, \pm 102$.
LO, NLO EW, NLO QCD cross sections are given in picobarns and compared
with corresponding values obtained with the aid of the program MCFM. Also
correction factors are shown in \%. The numbers illustrate good agreement
within statistical errors of MC integration.}
\end{table}
\begin{table}[!h]
\begin{center}
\begin{tabular}{|l|r|r|r|}
\hline
 pid             &  004        &  104         &   -104        \\ \hline
 LO              &  0.8291(1)  &  0.9277(1)   &   0.5883(1)   \\
 LO MCFM         &  0.8292(1)  &  0.9280(2)   &   0.5885(1)   \\
 NLO QCD         &  0.9685(3)  &  1.0897(3)   &   0.6866(3)   \\
 NLO MCFM        &  0.9686(1)  &  1.0901(2)   &   0.6870(1)   \\
$\delta_{QCD}$    &  16.81(3)   &  17.47(3)    &   16.72(5)    \\
 NLO EW          &  0.7877(1)  &  0.8672(2)   &   0.5508(1)   \\
$\delta_{EW}$     & -5.00(2)    & -6.52(2)     &  -6.38(3)     \\  \hline
\end{tabular}
\end{center}
\caption{Table 2:  The same is in Table 1 but for
processes of $HZ(W^{\pm})$ production, i.e. pid$ = 004, \pm 104$.}
\end{table}
\begin{table}[!h]
\begin{center}
\begin{tabular}{|l|r|r|}
\hline
pid              &   105               &   -105           \\ \hline
LO               &   5.134(1)          &   3.205(1)       \\
LO MCFM          &   5.133(1)          &   3.203(1)       \\
NLO QCD          &   6.921(2)          &   4.313(2)       \\           
NLO MCFM         &   6.923(2)          &   4.309(1)       \\
$\delta_{QCD}$    &   34.79(5)          &   34.56(8)       \\
NLO EW           &   5.022(1)          &   3.140(1)       \\ 
$\delta_{EW}$     &  -2.18(1)           &  -2.02(2)        \\  
\hline 
\hline
pid              &     106             &   -106           \\  \hline
LO               &   158.73(2)         &   95.18(2)       \\  
LO MCFM          &   158.69(7)         &   95.27(4)       \\
NLO QCD          &   152.13(9)         &   90.44(7)       \\     
NLO MCFM         &   152.07(14)        &   90.50(8)       \\
$\delta_{QCD}$    &    -4.17(6)         &   -4.08(8)       \\
NLO EW           &   164.44(5)         &   98.65(4)       \\ 
$\delta_{EW}$     &     3.59(3)         &    3.66(5)       \\  \hline
\end{tabular}
\end{center}
\caption{Table 3: The same is in Table 1 but for single top, s and t channels,
i.e. for pid$ = \pm 105,\pm 106$.}
\end{table}

$\bullet$ In Table 4 we show QCD and EW cross section contributions to the processes
of $HW^{\pm}$ production,
pid$ =\pm 104$, detailed over id's of the {\tt mcsanc} integrator. As is seen for
the chosen setup (Subsection 3.1) there
is strong cancellation of the Soft and Hard contributions {\tt id's=3,4} and gluon
induced contributions {\tt id's=5,6},
the sum of contributions {\tt id's=5+6} being negative. We remind that the sum of
all contributions is independent of the
unphysical parameters $\bar{\omega}$ and $m_q$.

\begin{table}[!h]
\begin{center}
\begin{tabular}{|c|c|r|r|}
\hline
QCD & pid       &   104        & -104          \\
\hline
    & id0       &   0.9277(1)  &  0.5883(1)    \\
    & id1       &   0.6916(1)  &  0.4860(1)     \\
    & id3       & -10.9233(1)  & -6.9139(1)    \\
    & id4       &  10.4547(1)  &  6.5737(1)    \\
    & id5       &  -0.9717(1)  & -0.6733(1)    \\
    & id6       &   0.9107(1)  &  0.6258(1)    \\
    & NLO QCD   &   1.0897(3)  &  0.6866(3)    \\
\hline
EW & pid        &   104        & -104          \\
\hline
   & id0        &  0.9277(1)   &  0.5884(1)    \\
   & id1        &  0.0100(1)   &  0.0070(0)    \\
   & id2        & -0.0560(0)   & -0.0349(0)    \\
   & id3        & -0.1592(1)   & -0.1003(0)    \\
   & id4        &  0.1448(1)   &  0.0907(1)    \\
   & NLO EW     &  0.8672(2)   &  0.5508(1)    \\ \hline
\end{tabular}
\end{center}
\caption{Table 4: EW and QCD radiative corrections in picobarns detailed over
{\tt id's} for {pid$ = \pm 104$}, parameter $\bar{\omega}=10^{-4}$.}
\end{table}

\clearpage

$\bullet$ The other comparisons.

1) EW corrections: a comparison of EW and QCD NLO corrections between {\tt mcsanc} and papers~\cite{Ciccolini:2003jy}
and ~\cite{Brein:2004ue} was done using corresponding setup.
We received good agreement within statistical errors with the results presented in the references.

2) QCD corrections: a comparison between inclusive LO and NLO cross sections from {\tt mcsanc}
and Table 1 of paper~\cite{Harris:2002md} was carried out using tuned setup.
Agreement within MC errors was found for LO, while for NLO only a qualitative agreement was reached, since
we did not manage to reproduce the corresponding value for $\alpha_s(r)$

More comparisons, including differential distributions (as in papers~\cite{Denner:2011id}, \cite{Denner:2011rn})
will be presented elsewhere.




\section{Conclusions}
\label{SectionConcl}

To match the experimental accuracy at the LHC, we direct our effort to developing
a programming environment for the calculation of processes at one loop level
and to creating the {\tt mcsanc} integrator with EW and QCD branches at hadron level.

In this paper we have presented results for EW and QCD corrections to the following processes:
$Z$ and $W$ production, $HZ$ and $HW^\pm$ production, and single top production processes.
Our investigation confirms that NLO precision level
and combination of EW and QCD corrections are mandatory
for the precision tag of the LHC experiments.

We thank  A.~Arbuzov, G.~Nanava and A.~Ochirov for essential contributions
to the {\tt mcsanc} project, and V.~Kolesnikov and R.~Sadykov
for useful discussions.

This work is partly supported by Russian Foundation for Basic Research grant
$N^{o}$ 10-02-01030-a.
A. Sapronov is cordially indebted to ``Dinastiya'' foundation 2011--2012.

\def\href#1#2{#2}
\addcontentsline{toc}{section}{MainRefInt}
\bibliographystyle{utphys_spires}
\bibliography{MainRefInt}

\providecommand{\href}[2]{#2}\begingroup\begin{thebibliography}{10}

\bibitem{Dittmaier:2012vm}
{LHC Higgs Cross Section Working Group} Collaboration, S.~Dittmaier,
  C.~Mariotti, G.~Passarino, R.~Tanaka, {\em et al.},
\href{http://www.arXiv.org/abs/1201.3084}{{\tt 1201.3084}}.

\bibitem{Dittmaier:2011ti}
{LHC Higgs Cross Section Working Group} Collaboration, S.~Dittmaier {\em et
  al.},
\href{http://www.arXiv.org/abs/1101.0593}{{\tt 1101.0593}}.

\bibitem{Balossini:2009sa}
G.~Balossini, G.~Montagna, C.~M. Carloni~Calame, M.~Moretti, O.~Nicrosini, {\em
  et al.}, {\em JHEP} {\bf 1001} (2010) 013,
\href{http://www.arXiv.org/abs/0907.0276}{{\tt 0907.0276}}.

\bibitem{Denner:2011id}
A.~Denner, S.~Dittmaier, S.~Kallweit, and A.~Muck, {\em JHEP} {\bf 1203} (2012)
  075,
\href{http://www.arXiv.org/abs/1112.5142}{{\tt 1112.5142}}.

\bibitem{Bernaciak:2012hj}
C.~Bernaciak and D.~Wackeroth, {\em Phys.Rev.} {\bf D85} (2012) 093003,
\href{http://www.arXiv.org/abs/1201.4804}{{\tt 1201.4804}}.

\bibitem{Barze:2012tt}
L.~Barze, G.~Montagna, P.~Nason, O.~Nicrosini, and F.~Piccinini, {\em JHEP}
  {\bf 1204} (2012) 037,
\href{http://www.arXiv.org/abs/1202.0465}{{\tt 1202.0465}}.

\bibitem{Yost:2012mf}
S.~Yost, V.~Halyo, M.~Hejna, and B.~Ward,
\href{http://www.arXiv.org/abs/1201.5906}{{\tt 1201.5906}}.

\bibitem{Andonov:2004hi}
A.~Andonov, A.~Arbuzov, D.~Bardin, S.~Bondarenko, P.~Christova, {\em et al.},
  {\em Comput.Phys.Commun.} {\bf 174} (2006) 481--517,
\href{http://www.arXiv.org/abs/hep-ph/0411186}{{\tt hep-ph/0411186}}.

\bibitem{Andonov:2008ga}
A.~Andonov, A.~Arbuzov, D.~Bardin, S.~Bondarenko, P.~Christova, {\em et al.},
  {\em Comput.Phys.Commun.} {\bf 181} (2010) 305--312,
\href{http://www.arXiv.org/abs/0812.4207}{{\tt 0812.4207}}.

\bibitem{Arbuzov:2005dd}
A.~Arbuzov, D.~Bardin, S.~Bondarenko, P.~Christova, L.~Kalinovskaya, {\em et
  al.}, {\em Eur.Phys.J.} {\bf C46} (2006) 407--412,
\href{http://www.arXiv.org/abs/hep-ph/0506110}{{\tt hep-ph/0506110}}.

\bibitem{Arbuzov:2007ke}
A.~Arbuzov, D.~Bardin, S.~Bondarenko, P.~Christova, L.~Kalinovskaya, {\em et
  al.}, {\em Eur.Phys.J.} {\bf C51} (2007) 585--591,
\href{http://www.arXiv.org/abs/hep-ph/0703043}{{\tt hep-ph/0703043}}.

\bibitem{Arbuzov:2007db}
A.~Arbuzov, D.~Bardin, S.~Bondarenko, P.~Christova, L.~Kalinovskaya, {\em et
  al.}, {\em Eur.Phys.J.} {\bf C54} (2008) 451--460,
\href{http://www.arXiv.org/abs/0711.0625}{{\tt 0711.0625}}.

\bibitem{Andonov:2009nn}
A.~Andonov, A.~Arbuzov, S.~Bondarenko, P.~Christova, V.~Kolesnikov, {\em et
  al.}, {\em Phys.Atom.Nucl.} {\bf 73} (2010) 1761--1769,
\href{http://www.arXiv.org/abs/0901.2785}{{\tt 0901.2785}}.

\bibitem{Bardin:2010mz}
D.~Bardin, S.~Bondarenko, L.~Kalinovskaya, V.~Kolesnikov, and W.~von Schlippe,
  {\em Eur.Phys.J.} {\bf C71} (2011) 1533,
\href{http://www.arXiv.org/abs/1008.1859}{{\tt 1008.1859}}.

\bibitem{Bardin:2011ti}
D.~Bardin, S.~Bondarenko, P.~Christova, L.~Kalinovskaya, V.~Kolesnikov, {\em et
  al.},
\href{http://www.arXiv.org/abs/1110.3622}{{\tt 1110.3622}}.

\bibitem{Campbell:arXiv1007.3492}
J.~M. Campbell and R.~K. Ellis,
  \href{http://www.arXiv.org/abs/arXiv/1007.3492}{{\tt arXiv/1007.3492}}.

\bibitem{MCFMhomepage}
{John Campbell, Keith Ellis, Ciaran Williams}, {MCFM homepage:
  http://mcfm.fnal.gov/}.

\bibitem{Ciccolini:2003jy}
M.~Ciccolini, S.~Dittmaier, and M.~Kramer, {\em Phys.Rev.} {\bf D68} (2003)
  073003,
\href{http://www.arXiv.org/abs/hep-ph/0306234}{{\tt hep-ph/0306234}}.

\bibitem{Brein:2004ue}
O.~Brein, M.~Ciccolini, S.~Dittmaier, A.~Djouadi, R.~Harlander, {\em et al.},
\href{http://www.arXiv.org/abs/hep-ph/0402003}{{\tt hep-ph/0402003}}.

\bibitem{Harris:2002md}
B.~Harris, E.~Laenen, L.~Phaf, Z.~Sullivan, and S.~Weinzierl, {\em Phys.Rev.}
  {\bf D66} (2002) 054024,
\href{http://www.arXiv.org/abs/hep-ph/0207055}{{\tt hep-ph/0207055}}.

\bibitem{Hahn:2004fe}
T.~Hahn, {\em Comput.Phys.Commun.} {\bf 168} (2005) 78--95,
\href{http://www.arXiv.org/abs/hep-ph/0404043}{{\tt hep-ph/0404043}}.

\bibitem{Hahn:CUBAhp}
T.~Hahn, {Cuba - a library for multidimensional numerical integration:
  www.feynarts.de/cuba/}.

\bibitem{Buttar:2008jx}
C.~Buttar, J.~D'Hondt, M.~Kramer, G.~Salam, M.~Wobisch, {\em et al.},
\href{http://www.arXiv.org/abs/0803.0678}{{\tt 0803.0678}}.

\bibitem{Dittmaier:2009cr}
S.~Dittmaier and M.~Huber, {\em JHEP} {\bf 1001} (2010) 060,
\href{http://www.arXiv.org/abs/0911.2329}{{\tt 0911.2329}}.

\bibitem{Sjostrand:2006za}
T.~Sjostrand, S.~Mrenna, and P.~Z. Skands, {\em JHEP} {\bf 0605} (2006) 026,
\href{http://www.arXiv.org/abs/hep-ph/0603175}{{\tt hep-ph/0603175}}.

\bibitem{Sjostrand:2007gs}
T.~Sjostrand, S.~Mrenna, and P.~Skands, {\em Comput. Phys. Commun.} {\bf 178}
  (2008) 852--867,
\href{http://www.arXiv.org/abs/0710.3820}{{\tt 0710.3820}}.

\bibitem{Barberio:1990ms}
E.~Barberio, B.~van Eijk, and Z.~Was, {\em Comput.Phys.Commun.} {\bf 66} (1991)
115--128.

\bibitem{Golonka:2005pn}
P.~Golonka and Z.~Was, {\em Eur.Phys.J.} {\bf C45} (2006) 97--107,
\href{http://www.arXiv.org/abs/hep-ph/0506026}{{\tt hep-ph/0506026}}.

\bibitem{Was:2008zu}
Z.~Was, P.~Golonka, and G.~Nanava, {\em Nucl.Phys.Proc.Suppl.} {\bf 181-182}
  (2008) 269--274,
\href{http://www.arXiv.org/abs/0807.2762}{{\tt 0807.2762}}.

\bibitem{Denner:2011rn}
A.~Denner, S.~Dittmaier, S.~Kallweit, and A.~Muck,
\href{http://www.arXiv.org/abs/1112.5258}{{\tt 1112.5258}}.

\end{thebibliography}\endgroup

\end{document}